\documentstyle[pra,aps,multicol,psfig]{revtex}
\begin{document}

\draft

\title{Quantum phases in an optical lattice}
\author{D. van Oosten,$^{1,2}$ P. van der Straten,$^2$ and
H. T. C. Stoof $^1$}
\address{$^1$Institute for Theoretical Physics,
         University of Utrecht, Princetonplein 5, \\
         3584 CC Utrecht,
         The Netherlands}
\address{$^2$Debye Institute,         
	 University of Utrecht, Princetonplein 5, \\
         3584 CC Utrecht,
         The Netherlands}

\maketitle

\begin{abstract}
We present the zero-temperature phase diagram of bosonic atoms in
an optical lattice, using two different mean-field approaches.
The phase diagram consists of various insulating phases and a superfluid
phase. We explore the nature of the insulating phase by calculating both
the quasiparticle and quasihole dispersion relation.
We also determine the parameters of our single band Bose-Hubbard model
in terms of the microscopic parameters of the atoms in the optical
lattice.
\end{abstract}


\begin{multicols}{2}
\section{Introduction}

Using the interference pattern of intersecting laser beams one can create
a periodic potential for atoms, which is known as an optical lattice
\cite{Grynberg,Jessen}.  Because one can confine atoms at separate
lattice sites, one can accurately control the interaction between the
atoms. This makes 
the optical lattice an important tool in spectroscopy, laser cooling
\cite{Chu} and quantum computing \cite{Brennen}. In the following we 
study some of the many-body aspects of such a lattice and in
particular Bose-Einstein condensation of atoms in an optical lattice. 
In contrast with the existing Bose-Einstein
condensation experiments in an harmonic trap, the quantum depletion
of the condensate in the case of Bose-Einstein condenstation in a lattice
can be very large. We can therefore expect interesting features.
 
If we assume that the atoms are cooled to within the lowest Bloch band of
the periodic potential, Jaksch {\it et al.} have shown that we can
describe the behavior of the atoms in an optical lattice with the
Bose-Hubbard hamiltonian

\begin{equation}
H=-t\sum_{\langle i,j\rangle} c^{\dagger}_i c_j +\frac{1}{2}U\sum_i
c^{\dagger}_ic^{\dagger}_i c_ic_i-\mu\sum_i c^{\dagger}_ic_i, 
\label{BoseHubbardH}
\end{equation} 
where the sum in the first term on the right-hand side is restricted to
nearest neighbours and $c^{\dagger}_i$ and $c_i$ are the creation and
annihilation operators of an atom at site $i$ respectively. The parameter
$t$ is the hopping 
parameter and $U$ is the interaction strength, which we always assume to
be positive in the following. The term involving the chemical potential
$\mu$ is added because we perform our calculations in the grand-canonical
ensemble. 

Qualitatively we expect that when there is an integer number
of particles at each site $i$ and $t\ll U$, the interaction between the
particles will make it energetically unfavourable for a particle to move
from one site to another. In this situation the gas is in what is known
as the Mott insulator phase \cite{Mott}. However, if we add in this
phase a particle to the system, this particle will only receive a small 
energetic penalty when it moves, because its interaction energy is the
same on each site.
For this reason, a gas with a non-integer number of bosons at
each site will be in a superfluid phase at zero-temperature. This
expectation has been shown to be correct using Quantum Monte Carlo
calculations \cite{Otterlo} and several mean-field approaches
\cite{Jaksch,Stroud,Sheshadri}. In particular, Ref.\ \cite{Jaksch}
numerically determines interesting features of cold bosonic atoms in an
inhomogeneous optical lattice. In this paper, we give a largely analytical
means of understanding the results obtained by these authors.

In order to describe the zero-temperature phase transition from the
superfluid to the  Mott-insulating phase analytically, we need to make
some appropriate mean-field approximation to the hamiltonian in 
Eq.\ (\ref{BoseHubbardH}). 
A more or less standard approach would be to
use the Bogoliubov approximation. In Sec. II we show that this
approximation does not predict the expected phase transition and we 
explain the absence of the phase transition. In Sec. III we
analytically investigate an alternative mean-field theory, proposed by
Sheshadri {\it et al.} \cite{Sheshadri}, and compare the
analytical results with exact numerical
results. In Sec. IV we discuss the properties of the Mott insulating
phases by calculating the quasiparticle and quasihole dispersions and
finally in Sec. V we relate the parameters $t$ and $U$ to experimental
parameters such as laser intensity and wavelength.

\section{Bogoliubov approximation}

We first transform the hamiltonian to momentum space by introducing 
creation and annihilation operators $a^\dagger_{\bf k}$ and $a_{\bf k}$
respectively, such that 

\begin{eqnarray}
c_i &=&  \frac{1}{\sqrt{N_s}} \sum_{\bf k}
a_{\bf k} e^{-i{\bf k}\cdot {\bf r}_i},\nonumber \\ & & 
\label{BFourier}
\\
c_i^{\dagger} &=&  \frac{1}{\sqrt{N_s}} \sum_{\bf k}
a_{\bf k}^{\dagger} e^{i{\bf k}\cdot {\bf r}_i}, \nonumber   
\end{eqnarray}
where $N_s$ is the number of lattice site and ${\bf r}_i$ is the
coordinate of site $i$. The wavevector ${\bf k}$ runs only over
the first Brillouin zone. For mathematical convenience we take only a
finite volume $V$, so that the momenta ${\hbar \bf k}$ are discretized,
which allows us to write sums instead of integrals in Eq.\
(\ref{BFourier}). Later we will take the continuum limit
$V\rightarrow\infty$. Using the fact that 
$\sum_i e^{-i({\bf k}-{\bf k}')\cdot {\bf r}_i}=
N_s \delta_{{\bf k},{\bf k}'}$, 
it is easily shown that the prefactor $1/\sqrt{N_s}$ ensures that the
total number of particles obeys $N=\sum_i c_i^{\dagger}c_i=
\sum_{\bf k} a_{\bf k}^{\dagger}a_{\bf k}$.

If we limit our description to cubic lattices with lattice distance $a$
and substitute Eq.\ (\ref{BFourier}) into the hamiltonian, we find

\begin{eqnarray}
H &=& \sum_{\bf k} 
\left( -\bar{\epsilon}_{\bf k}-\mu \right)  a_{\bf k}^{\dagger} a_{\bf k}
\nonumber \\
  &+& \frac{1}{2}\frac{U}{N_s} \sum_{\bf k} \sum_{{\bf k}'} \sum_{{\bf
k}''} \sum_{{\bf k}'''}   
a_{\bf k}^{\dagger} a_{{\bf k}'}^{\dagger} a_{{\bf
k}''}a_{{\bf k}'''} \delta_{{\bf k}+{\bf k}',{\bf k}''+{\bf k}'''},
\label{TussenHamil2}
\end{eqnarray}
where we defined  $\bar{\epsilon}_{\bf k}=2t\sum_{j=1}^d \cos(k_j a)$,
with $d$ the number of dimensions. 
For a Bose condensed gas the average number of condensate atoms $N_0$ is a
number much larger then one, which means that  $N_0=\langle a_{\bf
0}^{\dagger}a_{\bf 0} \rangle\approx \langle a_{\bf 0}a_{\bf 0}^{\dagger}
\rangle$ and we are allowed to take $N_0=\langle a_{\bf 0}^{\dagger} 
\rangle \langle a_{\bf 0} \rangle$. 

Since $\langle a_{\bf 0}^{\dagger}\rangle$ and $\langle a_{\bf 0} \rangle$
are complex conjugates, we conclude that $\langle a_{\bf
0}^{\dagger}\rangle =\langle a_{\bf 0} \rangle=\sqrt{N_0}$, where we have
chosen these expectation values to be real. The Bogoliubov approach 
consists of replacing the creation and annihilation operators by their
average $\sqrt{N_0}$ plus a fluctuation 

\begin{eqnarray}
a^{\dagger}_{\bf 0} &\rightarrow& \sqrt{N_0} + a^{\dagger}_{\bf 0},
\nonumber \\ && \label{Subsitutie1} \\
a_{\bf 0} &\rightarrow& \sqrt{N_0} + a_{\bf 0}, \nonumber
\end{eqnarray}
and minimizing the energy of the gas with respect to the number of
condensate atoms $N_0$.
At the minimum, the part of the hamiltonian that is linear in the
fluctuations must therefore be zero. Performing the above subsitution and
selecting the linear terms yields

\begin{equation}
H^{(1)}=(-\bar{\epsilon}_{\bf 0}-\mu+
\frac{U}{N_s} N_0 ) \sqrt{N_0}(a_{\bf 0}^{\dagger}+a_{\bf 0}),
\label{LinearPartHamil}
\end{equation}
where the superscript denotes the order in the
fluctuations. Since $H^{(1)}$ must be zero for all $a_{\bf 0}^{\dagger}$
and $a_{\bf 0}$ we conclude that in the lowest order approximation

\begin{equation}
\mu=Un_0 - zt,
\label{ChemPot}
\end{equation}
in terms of the condensate density $n_0=N_0/N_s$ and the number of
nearest neighbours $z=2d$. This expression can be easily understood since
the chemical potential is the energy needed to add one particle to the
system. Adding one particle results in an energy increase due to the
interaction with the $n_0$ particles already at each site, and an energy
decrease due to the possible hopping to one of $z$ nearest-neighbour
sites. 

Next we determine the effective hamiltonian $H^{\rm eff}$, which
contains only the parts of zeroth and second order in the fluctuations. 
The zeroth-order term is found by subtituting all creation and
annihilation operators by $\sqrt{N_0}$. 
To find the quadratic term, we substitute in the interaction term two
creation or annihilation operators at a time by $\sqrt{N_0}$ and write
down all possible combinations. Performing the summation over one of the
remaining momenta yields finally

\begin{eqnarray}
H^{\rm eff} &=& \left( 
-2z-\mu+\frac{1}{2} U n_0 \right) N_0 + \sum_{\bf k} 
\left( -\bar{\epsilon}_{\bf k}-\mu \right)  a_{\bf k}^{\dagger} a_{\bf k}
\nonumber \\
  &+& \frac{1}{2}Un_0 \sum_{\bf k} \left(
a_{\bf k}a_{-{\bf k}}+4a^{\dagger}_{\bf k}a_{\bf k} + a^{\dagger}_{-{\bf
k}}a^{\dagger}_{\bf k}
\right).
\label{TussenHamil3}
\end{eqnarray}
We can simplify this expression somewhat by using the commutation
relation $[a_{\bf k},a^{\dagger}_{\bf k}]=1$. If we also substitute Eq.\
(\ref{ChemPot}) and write $\epsilon_{\bf k}=2z-\bar{\epsilon}_{\bf k}$,
we find

\begin{eqnarray}
\label{EffectiveMatrixHamil}
H^{\rm eff} &=& -\frac{1}{2} U n_0 N_0 -
\frac{1}{2}\sum_{\bf k} \left( \epsilon_{\bf k} + U n_0 \right)\\
	&+&\frac{1}{2}\sum_{\bf k} \left(a_{\bf
k}^\dagger,a_{-{\bf k}} \right) \left[
\begin{tabular}{cc}
$\epsilon_{\bf k}+U n_0$       &$U n_0$ \\
$U n_0$  &$\epsilon_{\bf k}+U n_0$
\end{tabular}
\right]
\left(
\begin{tabular}{c}
$a_{\bf k}$ \\
$a_{-{\bf k}}^\dagger$
\end{tabular}
\right), \nonumber
\end{eqnarray} 
where the extra zeroth-order terms are generated by
the commutation of $a^{\dagger}_{\bf k}$ and $a_{\bf k}$.

The effective hamiltonian is diagonalized by a Bogoliubov
transformation. This implies that we define new creation and annihilation
operators $b^{\dagger}_{\bf k}$ and $b_{\bf k}$ for which the effective
hamiltonian is diagonal, by means of

\begin{equation}
\left(
\begin{tabular}{c}
$b_{\bf k}$\\   
$b_{-{\bf k}}^\dagger$
\end{tabular}
\right)=\left[
\begin{tabular}{cc}
$u_{\bf k}$   &$v_{\bf k}$\\
$v_{\bf k}^*$ &$u_{\bf k}^*$
\end{tabular} 
\right]\left(
\begin{tabular}{c}
$a_{\bf k}$\\
$a_{-{\bf k}}^\dagger$
\end{tabular}   
\right)
\equiv {\bf B}\left(
\begin{tabular}{c}
$a_{\bf k}$\\
$a_{-{\bf k}}^\dagger$
\end{tabular}   
\right).
\label{BogoliubovTransform}
\end{equation}    
To ensure that the operators $b^{\dagger}_{\bf k}$ and $b_{\bf k}$ still
obey the standard commutation relations for bosonic creation and
annihilation operators, we have to demand that the coefficients of matrix
${\bf B}$ obey

\begin{equation}
|u_{\bf k}|^2-|v_{\bf k}|^2=1.
\label{Normering}
\end{equation}
If we now substitute Eq.\ (\ref{BogoliubovTransform}) into
Eq.\ (\ref{EffectiveMatrixHamil}) and demand that the result reduces to
the diagonal hamiltonian

\begin{eqnarray}
H^{\rm eff}&=&-\frac{1}{2} U n_0 N_0 +\frac{1}{2}\sum_{\bf k}
\left[\hbar\omega_{\bf k}-\left( \epsilon_{\bf k} + U n_0
\right) \right]\nonumber \\
&+&\sum_{\bf k}\hbar
\omega_{\bf k} b_{\bf k}^\dagger b_{\bf k},
\label{EffectiveHamiltonian2}
\end{eqnarray}
we find that $u_{\bf k}$ and $v_{\bf k}$ must be solutions of the
following two equations:

\begin{eqnarray}
\left( (u_{\bf k})^2+(v_{\bf k})^2\right)Un_0
 - 2u_{\bf k}v_{\bf k}(\epsilon_{\bf k}+Un_0)
& = & 0, \nonumber \\ 
\label{U&Vvergelijking}
&& \\
\left(|u_{\bf k}|^2+|v_{\bf k}|\right)^2(\epsilon_{\bf k}+U
n_0) - (u_{\bf k}^* v_{\bf k}+u_{\bf k}
v_{\bf k}^*)Un_0 & = & \hbar\omega_{\bf k}.\nonumber
\end{eqnarray}
Using the normalization in  Eq.\ (\ref{Normering}), we can easily find
the solution

\begin{eqnarray}
\hbar\omega_{\bf k}
 & = &\sqrt{\epsilon_{\bf k}^2+2U n_0\epsilon_{\bf k}}, \nonumber \\
\label{UVergelijking}
&&\\
|v_{\bf k}|^2 & = & |u_{\bf k}|^2-1 = \frac{1}{2}\left(
\frac{\epsilon_{\bf k}+U n_0}{\hbar\omega_{\bf k}} -1
\right).\nonumber 
\end{eqnarray}

To also obtain the condensate density $n_0$, which until now has been
arbitrary, we now need to calculate the total density $n$ as given by our
effective hamiltonian. The total density is thus given by

\begin{equation}
n = \frac{1}{N_s}\sum_{\bf k}\langle a_{\bf k}^\dagger
a_{\bf k}
\rangle_{H^{\mbox{\tiny{eff}}}},
\label{DensityBasic}
\end{equation}
where the brackets  $\langle\rangle_{H^{\mbox{\tiny{eff}}}}$ denote
the expectation
value as calculated with the effective hamiltonian
$H^{\rm eff}$. For a Bose condensed gas, this density consists of two
parts: the density associated with the macroscopic occupation of the
one-particle ground-state, i.e., the condensate, and the density due to
the occupation of the higher lying one-particle states. In this case, the
condensate density equals the parameter $n_0$ and the density of the
non-condensate part is determined by an average over the quadratic
fluctuations, which will be a function of $n_0$.
Calculating the average over the quadratic fluctuations by means of Eq.\
(\ref{BogoliubovTransform}) yields first of all

\begin{eqnarray}
\label{Density1}
n  & = & n_0 \\
   & + & \frac{1}{N_s}\sum_{{\bf k}\neq0}\left[\left(|u_{\bf k}|^2
+|v_{\bf k}|^2
\right)  \langle b_{\bf k}^\dagger b_{\bf k}
\rangle_{H^{\mbox{\tiny{eff}}}}+|v_{\bf k}|^2    \right]. \nonumber
\end{eqnarray}
If we then use Eq.\ (\ref{UVergelijking}) and substitute the Bose
distribution evaluated at $\hbar\omega_{\bf k}$ for $ \langle b_{\bf
k}^\dagger b_{\bf k}\rangle_{H^{\rm eff}}$ we find that

\begin{eqnarray}
\label{Density2}
n  & = & n_0 \\
   & + &
\frac{1}{N_s}\sum_{{\bf k}\neq0}\left(\frac{\epsilon_{\bf k}+
U n_0}{\hbar
\omega_{\bf k}}\frac{1}{e^{\beta\hbar\omega_{\bf k}}-1}+
\frac{\epsilon_{\bf k}+U n_0-\hbar\omega_{\bf k}}{2\hbar
\omega_{\bf k}}\right). \nonumber 
\end{eqnarray}

In the zero-temperature limit, $\beta \rightarrow \infty$, the first
term in the summant is zero.
Taking the continuum limit by using $\sum_{\bf k} \rightarrow
V\int_{-\pi/a}^{\pi/a}d{\bf k}/(2\pi)^d$, changing
from momenta ${\bf k}$ to ${\bf q}=2\pi{\bf k}/a$, and realizing
that $N_s=V/a^d$, we arrive at the expression

\begin{equation}
n=n_0+\frac{1}{2}\int_{-1/2}^{1/2}d{\bf q} \left(\frac{\epsilon_{\bf q}+U
n_0}{\hbar\omega_{\bf q}}-1\right),
\label{DensityIntegral}
\end{equation}
with $\epsilon_{\bf q}=2t\sum_{j=1}^d
[1-\cos(2\pi q_j)]$ and $\hbar \omega_{\bf q}=(\epsilon_{\bf
q}^2+2 U n_0 \epsilon_{\bf q})^{1/2}$. We can now obtain the condensate
density by solving Eq.\ (\ref{DensityIntegral}) for $n_0$ for a fixed
value of $n$. We expect that at integer $n$, for a fixed value of $U/t$
there will be no superfluid solution and this will mark the phase
transition to the insulating phase as predicted by
\cite{Jaksch,Otterlo,Stroud,Sheshadri}. 

\subsection{Numerical results}

In Fig.\ \ref{BogoliubovPlaatje2D}(a) we plotted the result of this
calculation for a two dimensional lattice. We see from this figure,
that there is only a marginal difference between the case that $n=0.5$ and
$n=1.0$. In Fig.\ \ref{BogoliubovPlaatje2D}(b) we plotted the result for a
three dimensional lattice. In this case the difference between half
filling and integer filling is somewhat larger, but there is clearly no
critical value of  $U/t$ for which the condensate density goes to zero. 

These results lead to the suspicion that the phase transition to the
insulating phase is not present in this approximation. To verify this,
we investigate the limit of $U/t \rightarrow \infty$ in some detail.

\subsection{Asymptotic behavior}

When $U/t \rightarrow \infty$ we intuitively expect the system
to become an insulator, because it effectively means that the hopping
parameter goes to zero. We therefore expect that there are no
superfluid solutions as $U/t \rightarrow \infty$. We can see that in this
limit the integrand in the right hand side of Eq.\
(\ref{DensityIntegral}) behaves as  $(U n_0/2\epsilon_{\bf
q})^{1/2}$. One can also prove that $\epsilon_{\bf q} \leq
4\pi^2|{\bf q}|^2t$. This means that

\begin{equation}
\int_{-1/2}^{1/2}d{\bf q} \frac{\epsilon_{\bf q}+U
n_0 }{\sqrt{\epsilon_{\bf q}^2+2 U
\epsilon_{\bf q} n_0}} \geq 
\frac{1}{2\pi}\sqrt{\frac{U n_0}{2 t}}
\int_{-1/2}^{1/2}\frac{d{\bf q}}{|{\bf q}|}
\label{OnderGrensIntegraal}
\end{equation}
The integral at the right hand side of 
Eq.\ (\ref{OnderGrensIntegraal}) can be done
analytically in two dimensions and numerically in three dimensions. When
we call the result of the integration in $d$ dimensions $I_d$, we see that
Eq.\ (\ref{DensityIntegral}), for $U/t \rightarrow \infty$, reduces to

\begin{equation}
n\approx n_0+\frac{1}{4\pi}\sqrt{\frac{U n_0 }{2 t}} I_d -\frac{1}{2},   
\label{BogoLimitRootFind}
\end{equation}
where $I_2=2.22322$ and $I_3=2.38008$. This is a quadratic equation in
$\sqrt{n_0}$ which always yields a positive solution for $n_0$ given by

\begin{equation}
n_0=\left(
\frac{1}{2}\sqrt{\frac{{I_d}^2}{16\pi^2}\frac{U}{2 t}+4n+2}-
\frac{I_d}{8\pi}\sqrt{\frac{U}{2 t}}\right)^2.
\label{RootGevonden}
\end{equation}
We can correct for the error we made in Eq.\ 
(\ref{OnderGrensIntegraal}) by using a higher value for $I_d$, but while
this may change the value of $n_0$, it will still yield a positive
solution. 
We see from Eq.\ (\ref{RootGevonden}) that $\lim_{U/t\rightarrow\infty}
n_0 = 0$ as expected, so we can conclude that the Bogoliubov approximation
as described above does not predict the phase transition to the Mott
insulator phase in two and three dimensions. The reason for this is that
the Bogoliubov approach only approximately treats the interactions. 
As a result, the Bogoliubov approach cannot describe large depletions
of the condensate.
 
We also see from Eq.\ (\ref{OnderGrensIntegraal}) that in 
one dimension $I_1$ diverges. Substituting this in
Eq.\ (\ref{BogoLimitRootFind}), we see that there are no
Bose-condensed solutions ,i.e., solutions with $n_0\neq0$, in one
dimension. This is in accordance with the Mermin-Wagner-Hohenberg theorem
\cite{MWH1,MWH2,MWH3}.

As the Bogoliubov approximation fails to predict the phase transition to
the Mott insulator phase, we now consider a different mean-field theory
that treats the interactions exactly and approximates the kinetic energy
of the atoms in the optical lattice.

\section{Decoupling approximation}

To arrive at a mean-field approach, that is capable of
describing the Mott insulating phase, we start again from
Eq.\ (\ref{BoseHubbardH}). Analogous to the Bogoliubov approach, we
introduce the superfluid order parameter $\psi=\sqrt{n_i}=\langle
c^{\dagger}_i\rangle=\langle c_i \rangle$, where $n_i$ is the
expectation value of the number of particles on site $i$. Note that
we take the expectation values to be real, as before.  We now, however
construct a consistent mean-field theory by substituting

\begin{eqnarray}
c_i^{\dagger} c_j &=& \langle c_i^{\dagger} \rangle c_j + c_i^{\dagger}
\langle c_j \rangle  - \langle c_i^{\dagger} \rangle \langle c_j \rangle
\nonumber \\
&=&\psi\left( c_i^{\dagger}+ c_j  \right) -\psi^2,
\label{Substitutie}
\end{eqnarray}
into Eq.\ (\ref{BoseHubbardH}). Performing the substitution yields

\begin{eqnarray}
H^{\rm eff} &=& -zt \psi \sum_i \left( c_i^{\dagger} + c_i
\right) +zt\psi^2 N_s \nonumber \\
&+&
\frac{1}{2} U \sum_{i} c_i^{\dagger} c_i^{\dagger} c_i c_i- \mu \sum_{i}
c_i^{\dagger} c_i,
\label{BHHNaSubsitutie}
\end{eqnarray}
where $z=2d$ is again the number of nearest-neighbour sites and $N_s$ is
the total number of lattice sites, as before. This hamiltonian is diagonal
with respect to the site index $i$, so we can use an effective onsite
hamiltonian. If we introduce ${\bar U}=U/zt$, ${\bar \mu}=\mu/zt$ and the
number operator $\hat{n}_i=c^\dagger_i c_i$, we find

\begin{eqnarray}
H_i^{\rm eff} &=& \frac{1}{2}{\bar U}
\hat{n}_i\left(\hat{n}_i-1\right)-{\bar \mu} \hat{n}_i
-\psi\left( c^\dagger_i + c_i \right)+\psi^2,
\label{OnsiteHamil}
\end{eqnarray}
which is valid on each site $i$. We will therefore drop the subscript
$i$ in the following. Note that we scaled all the energies by a
factor $1/zt$, making this hamiltonian a dimensionless operator.

After writing Eq.\ (\ref{OnsiteHamil}) in matrix form with respect to an
occupation number basis, we can solve the problem numerically by
explicitely diagonalizing the part of the matrix with occupation number
below a certain maximum value \cite{Sheshadri}. Later we also follow this
procedure, but we first determine the phase diagram analytically using
second-order perturbation theory. 

\subsection{Second-order perturbation theory}

When we write $H^{\rm eff}=H^{(0)}+\psi V$, with

\begin{eqnarray}
H^{(0)}&=&\frac{1}{2}{\bar U} \hat{n}\left(\hat{n}-1\right)-{\bar \mu}
\hat{n}+\psi^2,\nonumber \\
\label{Splitsen}
&&\\
V&=&-\left( c^\dagger + c \right) \nonumber,
\end{eqnarray}
we see that in an occupation number basis the odd powers of the expansion
of the energy in $\psi$ will always be zero. If we denote the unperturbed
energy of the state with exactly $n$ particles by $E_n^{(0)}$, we find
that the unperturbed ground-state energy is given by

\[E_g^{(0)}= \left\{ E_n^{(0)} |n=0,1,2,...\right\}_{\rm min}.\]
Comparing $E_n^{(0)}$ and $E_{n+1}^{(0)}$ yields

\begin{equation}
E_g^{(0)}=\left\{ \begin{array}{ll}
         0 & \mbox{if ${\bar \mu}<0,$} \\
         \frac{1}{2}{\bar U} g(g-1)-{\bar \mu} g & \mbox{if
         ${\bar U} (g-1) < {\bar \mu} < {\bar U} g.$}
        \end{array} \right. 
\label{ClassicOplossingen}
\end{equation}

Next, we calculate the second-order correction to the energy with the 
well-known expression

\begin{equation}
E_g^{(2)}=\psi^2 \sum_{n \neq g} \frac{|\langle
g|V|n
\rangle|^2}{E_g^{(0)}-E_n^{(0)}},
\label{TweedeOrdeAlgemeen}
\end{equation}
where $|n\rangle$ denotes the unperturbed wave function with $n$
particles, of which the state with $n=g$ particles is the ground
state. Since the interaction $V$ couples only to states with one more or
one less atom than in the ground-state, we find

\begin{equation}
E_g^{(2)}=\frac{g}{{\bar U} (g-1)-{\bar \mu}}+\frac{g+1}{{\bar \mu}-{\bar
U} g}.
\label{TweedeOrdeAntwoord}
\end{equation}
If we now follow the usual Landau procedure for second-order phase
transitions by writing the ground-state energy as an expansion in $\psi$

\begin{equation}
E_g(\psi)=a_0(g,{\bar U},{\bar\mu})+a_2(g,{\bar
U},{\bar\mu})\psi^2+{\cal O}(\psi^4),
\label{TopTorde}
\end{equation}
and minimize it as a function of the superfluid order
parameter $\psi$, we find that $\psi=0$ when 
$a_2(g,{\bar U},{\bar \mu})>0$
and that $\psi\neq 0$ when $a_2(g,{\bar U},{\bar \mu})< 0$. 
This means that $a_2(g,{\bar U},{\bar \mu})=0$ signifies the boundary
between the superfluid and the insulator phases. Solving
\[a_2(g,{\bar U},{\bar \mu})=
\frac{g}{{\bar U}(g-1)-{\bar \mu}}+\frac{g+1}{{\bar \mu}-{\bar U}
g}+1=0,\] yields 

\begin{equation}
{\bar \mu}_{\pm}=\frac{1}{2} \left( {\bar U} (2g-1)-1 \right) 
\pm \frac{1}{2} \sqrt{{\bar U}^2-2{\bar U}(2g+1)+1},
\label{LobeVerg}
\end{equation}
where the subscript $\pm$ denotes the upper and lower halves of the Mott
insulating regions of phase space. Note that this result is exact within
our mean-field theory. Fig.\ \ref{PhaseDiagram} shows a plot
of Eq.\ (\ref{LobeVerg}) for $g=1,2,3$. By equating ${\bar \mu}_+$ and
${\bar \mu}_-$ we can find the point of smallest ${\bar U}$ for each
lobe. Denoting this critical value of ${\bar U}$ by ${\bar U}_c$ we have

\begin{equation} 
{\bar U}_c=2g+1+\sqrt{(2g+1)^2-1}.
\label{UcritVerg}
\end{equation}  
which yields ${\bar U}_c\approx 5.83$ for the $g=1$ insulator, a value
also found by \cite{Sheshadri}. 

\subsection{Fourth-order perturbation theory}
To find out more about the phase transition, we now carry out fourth-order
perturbation theory to find the rate with which the particle density 
increases as a function of ${\bar \mu}$. In the Appendix, we present a way
to calculate the higher-order terms in the perturbation series.
Using this procedure we can write the ground-state energy as

\begin{eqnarray}
E_g(\psi)&=&a_0(g,{\bar U},{\bar\mu})+
a_2(g,{\bar U},{\bar\mu})\psi^2\nonumber
\\ &+& a_4(g,{\bar U},{\bar \mu})\psi^4,
\label{TopVorde}
\end{eqnarray}
with

\begin{eqnarray}
a_4(g,{\bar U},{\bar\mu})
&=&\frac{g(g-1)}{\left({\bar U} (g-1)-{\bar\mu}\right)^2
                 \left({\bar U} (2g-3) - 2{\bar\mu}\right)
                }
                \nonumber \\
&+&   \frac{(g+1)(g+2)}{\left({\bar\mu}-{\bar U} g \right)^2
                \left(2{\bar\mu}-{\bar U}(2g+1)\right)
                }  
                \nonumber \\
        & - &   \left(\frac{g}{{\bar U} (g-1)-{\bar\mu}}+
                \frac{g+1}{{\bar\mu}-{\bar U} g} \right) \nonumber \\
        &\times&\left(
                \frac{g}{\left({\bar U} (g-1)-{\bar\mu}\right)^2}+   
                \frac{g+1}{\left({\bar\mu}-{\bar U} g\right)^2}\right).
\label{Vierdeorde}
\end{eqnarray}

In Figs.\ \ref{Energieverg}(a) and (b) we show plots of Eq.\
(\ref{TopVorde}) together with the result of an exact numerical
diagonalization of the effective hamiltonian.
As can be seen, the overlap is very good near the boundary given by 
Eq.\ (\ref{LobeVerg}). In Fig.\ \ref{Energiecusp} it can be seen that
the numerical result exhibits a
cusp when ${\bar U}={\bar\mu}$, which is not predicted by Eq.\ (\ref{TopVorde}).
This is due to the fact that in this particular case we need to use
first-order degenerate perturbation theory, because at 
${\bar\mu=n{\bar U}}$ the states with $n-1$ and $n$ particles per site form a
doubly degenerate ground-state. The resulting expression for the ground
state energy is now nonanalytic and given by

\begin{equation}
\left.E_g(\psi)\right|_{{\bar\mu=n{\bar U}}}=-\frac{1}{2}{\bar
U}n(n+1)+\psi^2-|\psi|\sqrt{n+1},
\label{CuspVerg}
\end{equation}
which is the solid line in Fig.\ \ref{Energiecusp}. Note that the
occurence of a cusp is analogous to the well-known Jahn-Teller effect in
solid-state physics \cite{Jahn}.

We now continue by calculating the average number of particles per site
in the grand-canonical ensemble by

\begin{eqnarray}
n&=&-\frac{\partial \langle H^{\rm eff} \rangle}{\partial \mu}
=-\frac{\partial E_g(\psi=\psi_{\rm min})}{\partial
{\bar \mu}}
\nonumber \\
&=&g-\frac{\partial}{\partial {\bar \mu}}\left( 
\frac{a_2(g,{\bar U},{\bar\mu})^2}
{4a_4(g,{\bar U},{\bar\mu})}
\right),
\label{DensityDus}
\end{eqnarray}
where 
$\psi_{\rm min}=[-a_2(g,{\bar U},{\bar\mu})/2a_4(g,{\bar
U},{\bar\mu})]^{1/2}$
is the minimum of Eq.\ (\ref{TopVorde}). Making use of the previous
results, we can now plot the density as a
function of ${\bar\mu}$ for a fixed value of ${\bar U}$. Between the edges
${\bar\mu}_{\pm}$, the density will remain constant because $\psi_{\rm
min}=0$ and the second term in the right-hand side of Eq.\
(\ref{DensityDus}) does not contribute. Outside that
region, the density will start to change with a nonzero slope.
In Fig.\ \ref{DensityProfile} this is plotted for ${\bar U}=11$. The
solid line shows the result of the calculation described above and the
dash-dotted line is a numerical result obtained by exact diagonalization.
As can be seen, the analytical results are in good agreement with the
numerical calculation. We can now also qualitatively understand the 
solution found numerically by Jaksch {\it et al.} \cite{Jaksch} for an
optical lattice in an external harmonic trap.
In a first approximation, we can describe the effect of a slowly varying
trapping potential by substituting ${\bar\mu}$ in Eq.\
(\ref{DensityDus})by ${\bar\mu}'={\bar\mu}+V({\bf r})$, where $V({\bf r})$
is the external trapping potential. Combining this with Fig.\
\ref{DensityProfile} yields the density profile found in
\cite{Jaksch}.

\section{Dispersion relations}

An important property of the Mott insulating phase is that the
fluctuation in the average number of particles per site goes to zero at
zero-temperature. Since these fluctuations can be described as  
quasiparticle and quasihole excitations, we will study these now.
We calculate the quasiparticle and quasihole dispersions using a
functional integral formalism. We start by deriving an expression for the
effective action. Readers unfamiliar with functional integrals may want
to skip to subsection IV B, where we discuss the results of the
calculation.

\subsection{The effective action}

We define complex functions $a_i^*(\tau)$ and $a_i(\tau)$, respectively,
and write the grand-canonical partition function as

\begin{eqnarray}
Z&=&\mbox{Tr} e^{-\beta\hat{H}}
=\int {\cal D} a^* {\cal D} a \exp{\left\{-S[a^*,a]/\hbar\right\} },
\label{Partitie1}
\end{eqnarray}
where the action $S[a^*,a]$ is given by

\begin{eqnarray} 
S[a^*,a] &=&\int_0^{\hbar\beta}d\tau \left[ \sum_i a_i^* \left(
\hbar \frac{\partial}{\partial\tau}-\mu \right) a_i
\right. \nonumber \\
          & - & \left. \sum_{ij} t_{ij} a_i^*a_j + \frac{1}{2} U \sum_i 
                a_i^* a_i^* a_i a_i  \right],
\label{Actie}
\end{eqnarray}
with $\beta=1/k_BT$, $k_B$ Boltzmann's constant and $T$ the
temperature. To decouple the hopping term, we perform a
Hubbard-Stratonovich transformation by adding a complete square to the
action, which then becomes

\begin{eqnarray}
\label{HSTransform}
S[a^*,a,\psi^*,\psi]&=&S[a^*,a]\\&+&\int_0^{\hbar\beta}d\tau
\sum_{ij}
\left( \psi_i^*-a_i^*\right) t_{ij} \left( \psi_j-a_j\right).\nonumber
\end{eqnarray}
Here $\psi^*$ and $\psi$ are the order parameter fields. To obtain an
effective action as a function of these fields, we rewrite Eq.\
(\ref{HSTransform}) as

\begin{eqnarray}
\label{Actie2}
&&S[a^*,a,\psi^*,\psi]    =      \int_0^{\hbar\beta}d\tau \left[
                                        \sum_i
                                a_i^*\left( \hbar
                                \frac{\partial}{\partial\tau}
                                -\mu \right) a_i
                                \right.\\ 
                         &+&  \frac{1}{2} U \sum_i a_i^* a_i^* a_i a_i
                                - \sum_{ij} t_{ij} \left( a_i^*
                                \psi_j + \psi_i^* a_j \right) 
                         +  \left. \sum_{ij} t_{ij} \psi_i^* \psi_j
                                \right],\nonumber
\end{eqnarray}
and integrate over the original fields $a_i^*$ and $a_i$. If we denote
by $S^{(0)}[a^*,a]$ the action for $t_{ij}=0$, we have explicitely that

\begin{eqnarray}
&&\exp{\left(-S^{\rm eff}[\psi^*,\psi]/\hbar\right)}\equiv
\exp{ \left(-\frac{1}{\hbar}\int_0^{\hbar\beta}d\tau \sum_{ij} t_{ij}
\psi_i^*\psi_j\right)} \nonumber \\
&\times&\int {\cal D} a^* {\cal D} a \exp{
\left\{-S^{(0)}[a^*,a]/\hbar \right\}}  \nonumber \\
&\times&\exp{ \left[ -\frac{1}{\hbar}\int_0^{\hbar\beta}d\tau
\left( - \sum_{ij} t_{ij} \left( a_i^* \psi_j + \psi_i^* a_j
\right) \right) \right]}.
\label{EffectiveActionIntegral}
\end{eqnarray} 

We can now calculate $S^{\rm eff}$ perturbatively by Taylor expanding the
exponent in the integrant of Eq.\ (\ref{EffectiveActionIntegral}) and
evaluating the various correlation functions of the field theory given by
$S^{(0)}$. This yields for the quadratic part of the effective action 

\end{multicols}
\widetext

\begin{eqnarray}
S^{(2)}[\psi^*,\psi]&=&-\frac{1}{2\hbar} \left<\left(
\int_0^{\hbar\beta}d\tau
\sum_{ij}t_{ij}\left(a_i^* \psi_j+\psi_i^*a_j\right)
\right)^2\right>_{S^{(0)}} + \int_0^{\hbar\beta}d\tau \sum_{ij} t_{ij}
\psi_i^*\psi_j
\nonumber\\
&=&-\frac{1}{2\hbar}\left<
\int_0^{\hbar\beta} \int_0^{\hbar\beta}d\tau d\tau'
 \sum_{iji'j'}   t_{ij} t_{i'j'}
\left(a_i^*\psi_j+\psi_i^*a_j\right)
\left(a_{i'}^*\psi_{j'}+\psi_{i'}^*a_{j'}\right)
\right>_{S^{(0)}}\nonumber \\
&+&\int_0^{\hbar\beta}d\tau \sum_{ij} t_{ij}\psi_i^*\psi_j.
\label{TweedeOrdeActie}
\end{eqnarray} 

\begin{multicols}{2}

\noindent If we perform the multiplication in the first term in the
right-hand side and use the information we have about the correlations in
the unperturbed system, i.e., 

\begin{eqnarray}
\left< a_i^* a_j^* \right>_{S^{(0)}} &=& \left< a_i a_j\right>_{S^{(0)}} =
0,
\nonumber\\
\left< a_i^* a_j\right>_{S^{(0)}} &=& \left< a_i a_j^*\right>_{S^{(0)}}=\left<
a_i a_i^*\right>_{S^{(0)}} \delta_{i,j},
\label{ClassicExpect}
\end{eqnarray}
we obtain in first instance

\begin{eqnarray}
\label{TweedeOrdeActie3}
&&S^{(2)}[\psi^*,\psi]=\int_0^{\hbar\beta}d\tau
\left\{ \sum_{ij} t_{ij}\psi_i^*(\tau)\psi_j(\tau)\right. \\ 
&-& \left.\frac{1}{\hbar}
\int_0^{\hbar\beta}d\tau' \sum_{iji'j'}t_{ij}t_{i'j'}
\psi_{j}^*(\tau) \left< a_i(\tau) a_{i'}^*(\tau')\right>_{S^{(0)}} \psi_{j'}
(\tau') \right\},\nonumber
\end{eqnarray}
where we have now shown the $\tau$ dependence of the fields explicitely 
for clarity reasons. Because we will only consider nearest-neighbour
hopping, we write

\begin{equation} 
t_{ij}=t_{ji}=\left\{
\begin{tabular}{cc}
$t$     & \mbox{for nearest neighbours}\\
$0$             & \mbox{otherwise.}
\end{tabular}
\right.
\label{Hoppingding}
\end{equation}  

First we treat the part of Eq.\ (\ref{TweedeOrdeActie3}) that is linear in
$t_{ij}$. We have

\begin{equation}
\sum_{ij} t_{ij}\psi_i^*(\tau)\psi_j(\tau) =\sum_i
t\psi_i^*(\tau)\psi_{i\pm\{1\}}(\tau),
\label{Lineairint}
\label{Springen}
\end{equation}
where $\pm\{1\}$ denotes all possible jumps to nearest
neighbours. In the case of one dimension this would simply be $\pm1$.
If we call the lattice spacing $a$ and introduce cartesian momentum
components $k_i$ with $i=1,\cdots ,d$, where $d$ is again the number of
dimensions, we find

\begin{equation}
\sum_{ij} t_{ij}\psi_i^*(\tau)\psi_j(\tau) = \sum_{\bf k}
2t\psi_{\bf k}(\tau)\psi^*_{\bf k}(\tau)\sum_{j=1}^d \cos{\left( k_j a
\right)} \label{Springen2}.
\end{equation}
Next we calculate the part that is quadratic in $t_{ij}$. We can treat
this part by looking at double jumps.  The expectation
value of  $\left< a_i a_{i'}^*\right>_{S^{(0)}}$ is proportional  
$\delta_{ii'}$ and independent of the site $i$, according to Eq.\
(\ref{ClassicExpect}). 
This means that we find, with similar notation as before, 

\begin{eqnarray}
&&\sum_{ji'j'}t_{ij}t_{i'j'} \psi_{j}^*(\tau) \left<
a_i(\tau) a_{i'}^*(\tau')\right>_{S^{(0)}} \psi_{j'} (\tau') \nonumber \\
&=&\left< a_i(\tau) a_{i}^*(\tau')\right>_{S^{(0)}} \sum_{jj'}
t_{ij}t_{ij'} \psi_{j}^*(\tau) \psi_{j'} (\tau') \nonumber \\
&=&t^2\left< a_i(\tau) a_{i}^*(\tau')\right>_{S^{(0)}} \sum_j\left\{
 z \psi_{j}^*(\tau) \psi_{j} (\tau')\right.\nonumber \\
&+&\left.\psi_{j}^*(\tau) \psi_{j\pm \{2\} } (\tau')
+\psi_{j}^*(\tau) \psi_{j\pm \{\sqrt{2}\} } (\tau')
\right\},
\label{DubbelSpringen}
\end{eqnarray}
with $z$ again the number of nearest neighbours. The first term in the
summant is a jump in each direction, followed by a jump back. The second
term indicates two jumps in the same direction and the third term is a
jump in each direction followed by a jump in a perpendicular
direction. Note that the third term is absent in one dimension. 
It can be shown that the complete double jump term reduces to

\begin{eqnarray}
&&\sum_{ji'j'}t_{ij}t_{i'j'} \psi_{j}^*(\tau) \left<
a_i(\tau) a_{i'}^*(\tau')\right>_{S^{(0)}} \psi_{j'} (\tau') \nonumber \\
&=&\left< a_i(\tau) a_{i}^*(\tau')\right>_{S^{(0)}}
\sum_{\bf k} \psi_{\bf k}^*(\tau)\psi_{\bf k}(\tau') (\bar{\epsilon}_{\bf
k})^2,
\label{DubbelSpringen2}
\end{eqnarray}
where we again used $\bar{\epsilon}_{\bf k}=2 t\sum_{j=1}^d \cos(k_j a)$.

To also treat the time dependence of the fields, we introduce Matsubara
frequencies  $\hbar\omega_n=\pi(2n)/\hbar\beta$ by

\begin{eqnarray}
\psi_{\bf k}(\tau)&=&\sum_n\frac{1}{\sqrt{\hbar\beta}} \psi_{{\bf k}n}
e^{-i\omega_n \tau}, \nonumber \\
\label{Matsubara}
&&\\
\psi^*_{\bf k}(\tau)&=&\sum_n\frac{1}{\sqrt{\hbar\beta}} \psi^*_{{\bf k}n}
e^{+i\omega_n \tau}.\nonumber 
\end{eqnarray}
To translate the expectation value of the fields into the
expectation value of operators, we introduce an (imaginary) time ordering
operator T. As a result

\begin{equation}
\left<a_i(\tau)a_{i'}^*(\tau')\right>_{S^{(0)}} =
\left<\mbox{T}\left[a_i(\tau)a_{i'}^{\dagger}(\tau')
\right]\right>_{S^{(0)}}.
\label{TijdOrdenen}
\end{equation}   
The time ordering can easily be expressed in Heavyside functions as

\begin{eqnarray}
\left<\mbox{T}\left[a_i(\tau)a_{i'}^{\dagger}(\tau')
\right]\right>_{S^{(0)}} &=&
\theta(\tau-\tau')\left<a_i(\tau)a_{i'}^{\dagger}(\tau')\right>_{S^{(0)}}
\nonumber \\
&+&
\theta(\tau'-\tau)\left<a_{i'}^{\dagger}(\tau')a_i(\tau)\right>_{S^{(0)}}.
\label{TijdOrdenen2}
\end{eqnarray}
If we use the unperturbed energies as given by (\ref{ClassicOplossingen}),
we thus find

\begin{eqnarray}
&&\left<a_i(\tau)a_{i'}^*(\tau')\right>_{S^{(0)}} 
\nonumber \\
& = & \theta(\tau-\tau')
(1+g) \exp{\left\{-
\left(E^{(0)}_{g+1}-E^{(0)}_{g} \right)(\tau-\tau')/\hbar\right\}}
\nonumber \\
&+& \theta(\tau'-\tau) g \exp{\left\{\left(
E^{(0)}_{g-1}-E^{(0)}_{g} \right)(\tau-\tau')/\hbar\right\}}.
\label{TijdOrdenen3}
\end{eqnarray}
Because $g$ minimizes $E_g^{(0)}$ we know that

\begin{eqnarray}
E^{(0)}_{g+1}-E^{(0)}_{g}& = & -\mu + g U > 0,
\nonumber \\
\label{Epsilonen}
&&\\
E^{(0)}_{g}-E^{(0)}_{g-1}& = & -\mu + (g-1) U < 0.\nonumber
\end{eqnarray}
Note that we use parameters $\mu$ and $U$ instead of ${\bar\mu}$ and
${\bar U}$, because we have not yet divided out the factor $zt$.
Combining the above with Eq.\ (\ref{TweedeOrdeActie3}) we find

\begin{eqnarray}
&&S^{(2)}[\psi^*,\psi]=\sum_n \sum_{\bf k}  |\psi_{{\bf k}n}|^2 \bar
\epsilon_{\bf k} \times\nonumber \\
&&\left(1-\frac{\bar \epsilon_{\bf k}}{\hbar} \int_{-\infty}^{0}d\tau'
(1+g) \exp{\left\{\left(-i\hbar\omega_n-\mu+gU\right) \tau'/\hbar\right\}}
\right. \nonumber \\
 &&-\left. \frac{\bar \epsilon_{\bf k}}{\hbar} \int_{0}^{\infty}d\tau'
g\exp{\left\{-\left(i\hbar\omega_n+\mu-(g-1)U\right)
\tau'/\hbar\right\}}\right). \nonumber \\
\label{Actievoordispersie}
\end{eqnarray} 
Performing the $\tau'$ integration we then easily obtain

\begin{eqnarray}
\label{ActievoordispersieInter}
&&S^{(2)}[\psi^*,\psi]=\sum_n \sum_{\bf k}|\psi_{{\bf k}n}|^2 (\bar
\epsilon_{\bf k})\times \\
&& \left[ 1-(\bar \epsilon_{\bf k})\left(\frac{g+1}{-i\hbar
\omega_n-\mu+gU}+
\frac{g}{i\hbar\omega_n+\mu-(g-1)U} \right)\right]. \nonumber
\end{eqnarray}
Note that this result is exact within our mean-field theory. It contains
all powers of the frequencies and momenta and no gradient expansion
has been applied. This is important because the elementary excitation are
gapped as we will show in the next section.

We can find an equation for real energies $\hbar \omega$ by subtituting
$i\omega_n \rightarrow\omega$ and equating the remaining factor to
zero. This gives

\begin{equation}
0=\left[ 1-(\bar \epsilon_{\bf k})
\left(\frac{g+1}{-\hbar \omega-\mu+gU}+
\frac{g}{\hbar\omega+\mu-(g-1)U} \right)\right].
\label{MatsubaraDispersie}
\end{equation}
Ultimately this yields the result Eq.\ (\ref{DispersieExpliciet})
given below in subsection IV B.

\subsection{Results}

Now we will explore the results of the calculation presented in the
previous subsection. The quasiparticle and quasihole dispersions are given
by

\begin{eqnarray} 
\hbar
\omega_{qp,qh}&=&-\mu+\frac{U}{2}(2g-1)
-\frac{\bar \epsilon_{\bf k}}{2}\nonumber \\&\pm&
\frac{1}{2}\sqrt{(\bar \epsilon_{\bf k})^2-(4g+2)U{\bar \epsilon}_{\bf 
k}+U^2}.
\label{DispersieExpliciet}
\end{eqnarray}
In Fig.\ \ref{PHDispersie}(a) we show for ${\bf k}=0$ a plot of the above
equations. The dotted lines indicate the asymptotes of Eq.\
(\ref{DispersieExpliciet}), which are given by

\begin{eqnarray}
\lim_{U\rightarrow\infty} \hbar \omega_{qp} &=&
-\mu+gU-(g+1){\bar \epsilon}_{\bf 0}
\nonumber\\
&=&E_{g+1}^{(0)}-E_{g}^{(0)}-(g+1)zt,
\nonumber\\
\lim_{U\rightarrow\infty} \hbar \omega_{qh} &=&
-\mu+(g-1)U+g{\bar \epsilon}_{\bf 0} 
\nonumber \\
&=&E_{g}^{(0)}-E_{g-1}^{(0)}+gzt,
\label{DispersieAsymptote}
\end{eqnarray}
with $E_{g+1}^{(0)}-E_{g}^{(0)}$ and $E_{g}^{(0)}-E_{g-1}^{(0)}$ given by
Eq.\ (\ref{Epsilonen}). The difference between 
Eq.\ (\ref{DispersieAsymptote}) and  Eq.\ (\ref{Epsilonen}) is caused by
the fact that Eq.\ (\ref{Epsilonen}) is calculated for $t=0$. It can
easily be understood that for $t\neq0$, the first-order correction is due
to the hopping terms $c^{\dagger}_j c_i t$, where site $j$ is one of the
nearest neighbours of site $i$. When we have $g$ particles in all lattice
sites and we add one particle to site $i$, we have $\langle c^{\dagger}_j
c_i \rangle = g+1$, so the effective hopping parameter for a particle is
$(g+1)t$. However, when we remove a particle from site $i$, we
have $\langle c^{\dagger}_i c_j \rangle = g$, which represents a particle
hopping to site $i$ from one of its nearest neighbours. The effective
hopping parameter for a hole is therefore only $gt$.
In combination, we see that in the limit of $U\rightarrow\infty$, Eq.\
(\ref{DispersieExpliciet}) indeed reduces to a physically intuitive
result.

As shown above, the slopes of the asymptotes
differ exactly by $U$, so in the limit of $U/zt\rightarrow\infty$ the gap
for the creation of a quasiparticle-quasihole pair is equal to $U$.
We can find a first approximation for the dispersion of the density
fluctuations by substracting the two solutions, which yields

\begin{eqnarray}
\epsilon_{\bf k}&=&\hbar \omega_{qp}-\hbar \omega_{qh}\nonumber\\ &=&
\sqrt{(\bar \epsilon_{\bf k})^2
-(4g+2)U\bar \epsilon_{\bf k}+U^2}.
\label{ParticleHoleDispersie}
\end{eqnarray}
In Fig.\ \ref{PHDispersie}(b) we show again for ${\bf k}=0$ a plot of the
above equation as a function of ${\bar U}=U/zt$ for $g=1$.
We can see that there is a band gap, which proves that the MI phase is
indeed an insulator and we also see that the band gap disappears as we
approach the critical value ${\bar U}_c=U_c/zt\approx 5.83$ that was found
earlier. For smaller values of ${\bar U}$ we are in the superfluid phase,
which according to the Hugenholtz-Pines theorem is expected to always have
gapless density fluctuations.

\section{Microscopic parameters}

To estimate the experimental feasibility of observing the described phase
transition, we now relate the parameters $t$ and $U$ to the
microscopic parameters. Because we have an experimental interest in
sodium,  we will calculate these parameters for sodium atoms trapped in a
lattice constructed with four laser beams. 
To calculate the hopping parameter $t$, we
calculate the overlap between single particle ground-state wave functions
in neighboring sites. To calculate the interaction strength $U$, we use
the pseudopotential method. 

First we calculate the optical potential experienced by the
atoms, following the approach of Petsas {\it et al.} \cite{Petsas}. 
We describe a $J=1/2 \rightarrow J=3/2$ transition and 
choose a laser beam configuration with two pairs of
laser beams. Each pair lies in a plane and the planes are perpendicular to
each other. All beams have the same angle $\theta$ with respect to the
intersection of the two planes. We choose the
quantization axis along the intersection and label it as the 
$z$-axis. Furthermore, we choose the
polarization of the laser beams linear and perpendicular to the plane
spanned by the pairs of beams. The configuration is illustrated in Fig.\
\ref{Configuratie}. It should be noted that it is also possible to simply
superimpose $d$ standing waves to obtain a $d$ dimensional lattice,
but this requires stabilization of the relative phases of the laser beams. 
We define $I_b$ as the sum of the intensity of the beams and if $k$ is the
magnitude of the ${\bf k}$-vector, we define $k_{\perp}=k\sin{\theta}$ and
$k_{//}=k\cos{\theta}$. If we add the electric field components and
express them in spherical components

\begin{eqnarray}
E_+&=&\frac{-1}{\sqrt{2}}(E_x-iE_y),\nonumber\\
E_-&=&\frac{1}{\sqrt{2}}(E_x+iE_y),\nonumber\\
E_0&=&E_z,
\label{BerekenSpherischeBijdrage}
\end{eqnarray}
we find that the spatial dependence of the intensity of the resulting
light field is given by

\begin{eqnarray}
I_{\pm}/I_b&=&\frac{1}{2}\left(
\cos^2{(k_{\perp} x)}+\cos^2{(k_{\perp} y)}\right.\nonumber\\
&\pm&\left. 2\cos{(k_{\perp}
x)}\cos{(k_{\perp} y)}\cos{(2k_{//}
z)}
\right), \nonumber\\
I_0/I_b&=&0.
\label{Intens3d1}
\end{eqnarray}
Note that at the minima of $I_{\pm}$ the polarization is purely
$\sigma^{\pm}$. Also note that since the linear component is always zero,
the two ground-state levels are not coupled. 

Following Nienhuis {\it et al.} \cite{Nienhuis} we can now calculated the
optical potential. Because of the fact that the ground-states
are not
coupled, we can treat them seperately. With $\delta$ the detuning,
$\Gamma$ the rate of spontaneous decay and $\Omega_{\pm}$ the Rabi
frequencies for the $\sigma^{\pm}$ components of the light field, we can
write the potential for the $m_j=\pm 1/2$ level in the limit of low
saturation as:

\begin{equation}
V_\pm=\frac{1}{2}\frac{\hbar \delta}{1+4(\delta/\Gamma)^2}\left[
\frac{2|\Omega_\pm|^2}{\Gamma^2}+\frac{1}{3}\frac{2|\Omega_\mp|^2}
{\Gamma^2}
\right],
\label{AppPotential1}
\end{equation}
where the factor $\frac{1}{3}$ arises because of the Clebsch-Gordan
coefficients for $J=1/2 \rightarrow J=3/2$ transitions. 

Now we define a convenient prefactor: 

\begin{equation}
V_b=\frac{1}{2}\frac{\hbar \delta
s_0}{1+4(\delta/\Gamma)^2}
=\frac{\hbar \delta s}{2},
\label{LightShiftOneBeam}  
\end{equation}
where $s=s_0/\left(1+4(\delta/\Gamma)^2 \right)$ is the off-resonance saturation
parameter and $s_0=2|\Omega|^2/\Gamma^2$ is the on-resonance saturation parameter,
which is usually written as $s_0=I/I_s$. The saturation intensity $I_s$
is a constant for a given transition.
If we substitute Eq.\ (\ref{LightShiftOneBeam}) in 
Eq.\ (\ref{AppPotential1}), we
find

\begin{eqnarray}
V_\pm&=&V_b \left( \frac{I_\pm}{I_b} + \frac{1}{3}\frac{I_\mp}{I_b}
\right).\nonumber\\
&=&\frac{2}{3}V_b\left( \cos^2{(k_\perp x)} + \cos^2{(k_\perp y)}
\right.\nonumber \\
&\pm&\left.
\cos{(k_\perp x)}\cos{(k_\perp y)} \cos{(2k_{//}z)}
\right).
\label{AppPotential2}
\end{eqnarray}

We now write the hamiltonian of a particle in the potential as:

\begin{equation}
H_{\rm opt}=\frac{p^2}{2m}+V_\pm,
\label{OpticalHamiltonian}
\end{equation}
and solve the time-independent Schr\"odinger equation variationally by 
assuming an isotropic gaussian wave function and minimizing the energy as 
a function of width of the gaussian. If we call $\beta$ the width of the
wave function, we can write the normalized wave function as

\begin{equation}
\Psi ({\bf r}) = \langle {\bf r} | \Psi \rangle =
\left( \frac{1}{\pi \beta^2 }\right)^{3/4}
e^{-|{\bf r}|^2/2\beta^2}. 
\label{Trialfunction}
\end{equation}
We assume we have a spin polarized
sample of atoms, so we can use either the $V_+$ or the $V_-$ potential.

For simplicity we now calculate the parameters for a one dimensional
lattice. For the lattice configuration in Fig.\ \ref{Configuratie} this
gives approximate results, but for a phase stabilized superposition of three
standing waves, the results are immediately applicable.
In this case, the potential reduces to
$V_\pm=2V_b[2\pm\cos{(2kz)}]/3+\kappa_\perp (x^2+y^2)/2$, where the
transverse potential is caused by the fact that the laser beam has a finite
width.  
If we assume the wavefunction is tightly localized in the center of the
local potential well, we can approximate the potential as an harmonic
potential $V_+=2V_b+\kappa {\bf r}^2/2$ with $\kappa=-8V_bk^2/3$, where
we assume we can adjust the width of the laser such that
$\kappa_\perp\approx\kappa$. These approximations yield the well-known
equations for the width and the level splitting in the potential

\begin{eqnarray}
\beta=\left( \frac{\hbar^2}{m\kappa}\right)^{1/4},
\hbar\omega=\sqrt{\kappa/m}.
\label{Harmonic}
\end{eqnarray}

Using the above width $\beta$ we calculate the value of the interaction
strength $U$ with the pseudopotential method. According to
\cite{Tiesinga} this is valid for sodium even if the
width of the trapping volume is of the order of the scattering
length. In general the interaction strength between two atoms in the
same one particle wave function is given by

\begin{equation}
U=\int d{\bf r} \int d{\bf r}' \Psi^*({\bf r}) \Psi^*({\bf r}')
V_{\rm int}({\bf r}-{\bf r}')  \Psi({\bf r}) \Psi({\bf
r}'),
\label{InteractionStrength}
\end{equation}
where $V_{\rm int}({\bf r}-{\bf r}')$ is the interaction
potential. If we approximate the potential as

\begin{equation}
V_{\rm int}({\bf r}-{\bf r}')=\frac{4\pi
a_s\hbar^2}{m}\delta({\bf r}-{\bf r}'),
\label{InteractionPotential}
\end{equation}
we can write Eq.\ (\ref{InteractionStrength}) as:
\begin{eqnarray}
U&=&\frac{4\pi a_s \hbar^2}{m} \int d{\bf r}
\Psi^*({\bf r}) \Psi^*({\bf r}) \Psi({\bf r})\Psi({\bf r})
\nonumber \\
&=&
\frac{4\pi a_s \hbar^2}{m}  \int d{\bf r}|\Psi({\bf r})|^4
=\frac{4\pi a_s \hbar^2}{m \beta^3 \pi^{3/2}}\nonumber \\
&=&\frac{2\hbar\omega}{\sqrt{2\pi}}\left(\frac{a_s}{\beta}\right),
\label{SimpelInteractionStrength}
\end{eqnarray}
where $a_s$ is the triplet s-wave scattering length. According to
\cite{Scattering} the value of the scattering length for a
spin polarized sodium-sodium collision is $a_s=(85\pm3)a_0$.
Note that the use of a one-band model is justified when
$U\ll\hbar\omega$, or $\beta\gg 2 a_s/(2\pi)^{1/2}\approx 3.5 \rm
nm$.

Next we calculate the value of the hopping parameter $t$. In the tight
binding limit $t$ is given by

\begin{equation}
t=-\int d{\bf r}
\Psi^*({\bf r})\left(\frac{p^2}{2m}+V_{\pm} \right)
\Psi({\bf r}+a\hat{e}_j),
\label{HoppingIntegral}
\end{equation}
where $\hat{e}_j$ is an axis vector along a lattice direction, so that
when $\Psi({\bf r})$ is the ground-state wave function,
$\Psi({\bf r}+a\hat{e}_j)$ is the ground-state wave function of an atom
at a neighboring site. One can show that product of two
wavefunctions at neighbouring sites is a gaussian function centered
around ${\bf r}+a\hat{e}_j/2$. We can therefore approximate the
potential around the maximum of the barrier by:
$V_\pm=2V_b/3\mp\kappa{\bf r}^2/2$. Substituting this into Eq.\
(\ref{HoppingIntegral}) ultimately yields

\begin{equation}
t=\frac{\hbar\omega}{8}\left[ 1-\left( \frac{2}{\pi}\right)^2\right]\left(
\frac{a}{\beta}\right)^2e^{-\frac{1}{4}\left(a/\beta\right)^2}
\label{HoppingValue}
\end{equation}

Figs.\ \ref{Parameterplaatjes}(a) and (b) show plots of $U/E_r$ and
$t/E_r$ respectively, with $E_r=\hbar^2k^2/2m$ the recoil energy. Both are
plotted as a function of the trap depth $V_{\rm trap}=-4V_b/3$.
The values were calculated for a laser wavelength
of $600$ nm. The saturation parameter needed to reach these trap
depths is in the order of $10^5$, which is not unrealistic experimentally.

\noindent Fig.\ \ref{Uovertplaatje} shows also $U/zt$ as a function of the
trap depth, for a wavelength of $600$ nm, in one, two and three
dimensions. Again, the saturation parameter is in the order of $10^5$.
As can been seen, the desired critical value is reached in all three
dimensions.
The value of the width $\beta$ lies between $12\%$ and $8\%$ of the
wavelength in the range considered in the above plots. This implies that
both the harmonic approximation and the use of the one-band model are
justified.

\section{Conclusion}

Due to the absence of the superfluid-insulator phase transition in the
Bogoliubov approach, we conclude that the interaction is the dominant
component in this phase transition. When the interaction energy is treated
exactly, the theory indeed predicts a phase transition. The mean-field
theory predicts a phase transition even in one dimension, 
which we expect to survive as a Kosterlitz-Thouless transition when
fluctuations are incorperated. However a definite prove of this requires
further study. 

We analytically calculated
the phase diagram and the particle and hole dispersion relations in the
insulator phase.  A first-order approximation to the dispersion of the 
density fluctuations shows that the system indeed goes from a gapped
to a gapless phase. A calculation of this dispersion below
the critical value for $U/zt$ will have to be done in order to check the
presence of linear dispersion that would verify the assumption that the
phase with $\psi\neq0$ is indeed superfluid. The one-band model we used to
calculate the parameters for sodium gives optimistic results for
future experiments, within the range of parameters it allows. 

\appendix{
\section*{The perturbation series}

A powerful approach to calculating higher-order terms in the perturbation
series  is derived in \cite{Messiah}. Here we only give the result of that
derivation.
If we denote by $|n\rangle$ the unperturbed wave functions and $E_n^{(0)}$
the unperturbed energies, we can define an operator

\begin{equation}
S^k_a=\left\{ \begin{array}{ll}
                -|a\rangle \langle a| & \mbox{if $k=0$} \\
                \sum_{n\neq a} \frac{|n\rangle \langle n
|}{\left(E_a^{(0)}-E_n^{(0)}\right)^{k+1}} & \mbox{if $k>0$}
        \end{array} \right.,
\label{Notatie} 
\end{equation}
and one can prove that the $n$-th order correction on the energy
$E_a^{(0)}$ is given by

\begin{eqnarray}
E_a^{(n)}& = & \mbox{Tr} \left[ \sum_{\{n-1\}} S^{k_0}_a
\hat{V}...\hat{V} S^{k_n}_a\right],
\label{Eallorders}
\end{eqnarray}
where ${\{n\}}=\{k_0,...,k_n|k_0+...+k_n=n\}$.
In the case of $n=2$, this quickly gives the well-known result

\begin{eqnarray}
E_a^{(2)}& = & \mbox{Tr} \left[\sum_{\{1\}} S^{k_0}_a \hat{V} S^{k_1}_a
                \hat{V} S^{k_2}_a\right]\nonumber \\
        & = &  \langle a |  \hat{V} S^1_a \hat{V}
                | a \rangle, \nonumber \\
	& = & \sum_{n \neq a} 
                \frac{\left|\langle n|\hat{V}|a\rangle\right|^2}
                {\left(E_a^{(0)}-E_n^{(0)}\right)}.
\label{Eantwee}
\end{eqnarray}
The same can be done for $E_a^{(1)}$, $E_a^{(3)}$ and $E_a^{(4)}$. The first two
can be shown to involve only terms proportional to odd orders of $V$ and
with $V\propto(c^{\dagger}+c)$ these are of course zero. The fourth-order
term is in general given by

\begin{eqnarray}
&&E_a^{(4)}= \mbox{Tr} \left[\sum_{\{3\}} S^{k_0}_a \hat{V} S^{k_1}_a
                \hat{V} S^{k_2}_a \hat{V} S^{k_3}_a  \hat{V} S^{k_4}_a 
		\right]
		\nonumber \\
	&& =   \langle a |  \hat{V} S^1_a \hat{V} S^1_a
                \hat{V} S^1_a \hat{V} | a \rangle -
                \langle a |  \hat{V} S^1_a \hat{V} | a \rangle
                \langle a |  \hat{V} S^2_a \hat{V} | a \rangle
                \nonumber \\
        && -    2 \langle a |  \hat{V} | a \rangle \langle a |  \hat{V}
                S^1_a \hat{V} S^2_a \hat{V} | a \rangle
                + \langle a |  \hat{V} | a \rangle^2
                \langle a |  \hat{V} S^3_a \hat{V} | a \rangle.\nonumber\\
\label{Eanvier}
\end{eqnarray}
If we drop the terms containing expectation values of odd powers of
$V$ and substitute  Eq.\ (\ref{Notatie}), we find
}

\begin{eqnarray}
&&E_a^{(4)}  =  \langle a |  \hat{V} S^1_a \hat{V} S^1_a
                \hat{V} S^1_a \hat{V} | a \rangle -
                \langle a |  \hat{V} S^1_a \hat{V} | a \rangle
                \langle a |  \hat{V} S^2_a \hat{V} | a \rangle
                \nonumber \\
         &&  = \sum_{n,p,q \neq a}  \langle a|\hat{V}|n\rangle\left(-E_a^{(2)}
                \frac{\langle n|\hat{V}|a\rangle} 
		{\left(E_a^{(0)}-E_n^{(0)}\right)^{2}}\right. \nonumber \\
         &&  +  \left.\frac{\langle n | \hat{V} | p \rangle}
                {\left(E_a^{(0)}-E_n^{(0)}\right)} 
                \frac{\langle p |  \hat{V} | q \rangle}
                {\left(E_a^{(0)}-E_p^{(0)}\right)}
                \frac{\langle q |  \hat{V} | a \rangle}
                {\left(E_a^{(0)}-E_q^{(0)}\right)} \right),\nonumber \\
\label{Eatvier}
\end{eqnarray}  
which we have used to derive Eq.\ (\ref{Vierdeorde}).

\figure{
\psfig{figure=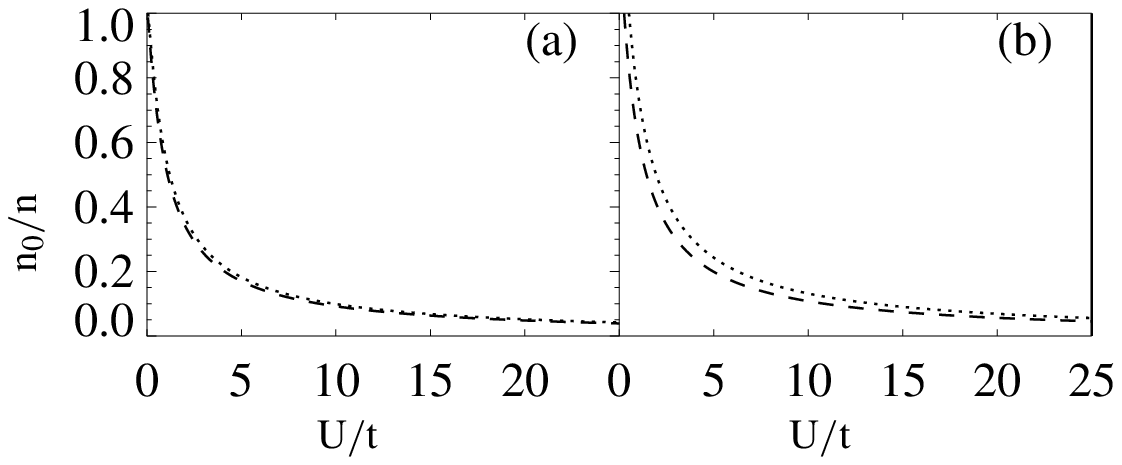,width=8cm}
\caption{\narrowtext The condensate fraction $n_0/n$ (a) in a 
two-dimensional optical lattice and (b) a three-dimensional optical
lattice, both as a function of $U/t$ for $n=0.5$ (dashed line) and
$n=1.0$ (dotted line).}
\label{BogoliubovPlaatje2D}}

\figure{
\psfig{figure=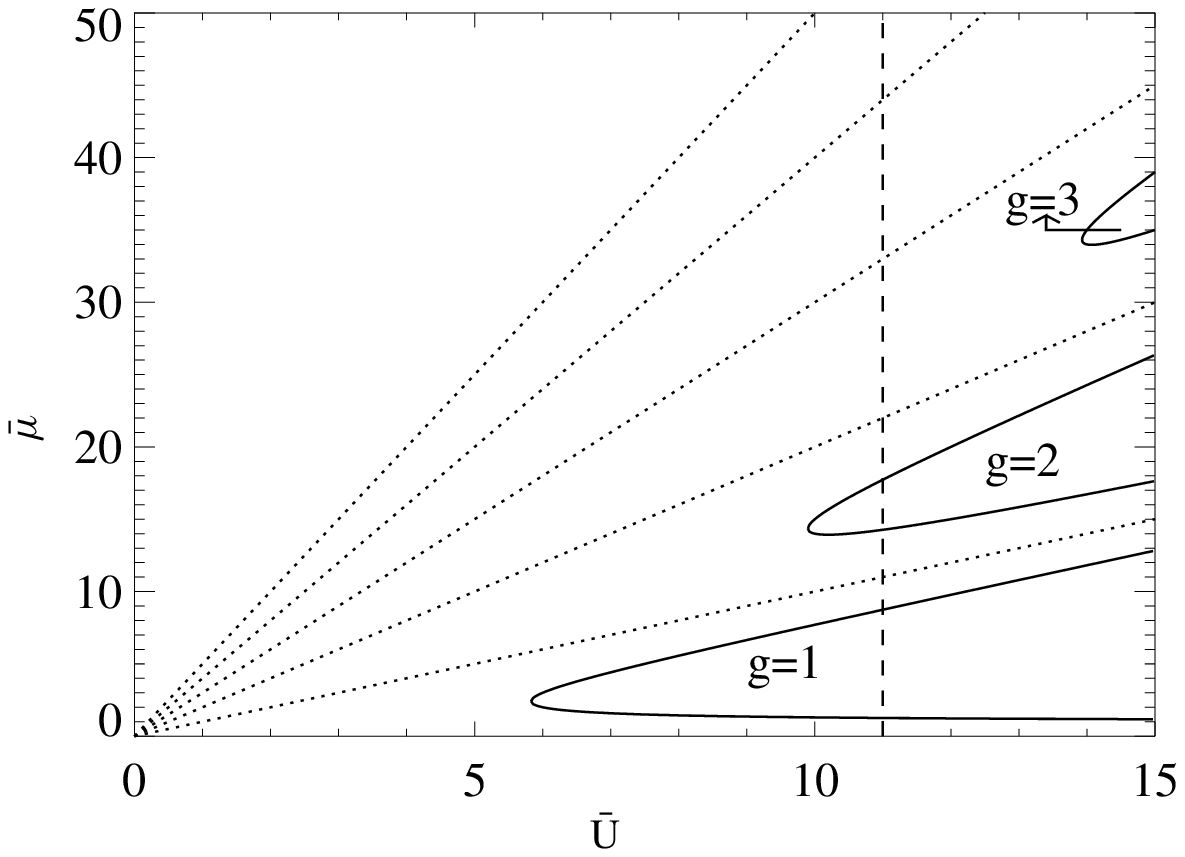,width=8cm}
\caption{\narrowtext Phase diagram of the Bose-Hubbard Hamiltonian
as obtained from second-order perturbation theory (solid lines). The
dotted lines indicate the zeroth-order phase diagram.
Later on, Fig.\ \ref{DensityProfile} is taken along the dashed line in
this figure.}
\label{PhaseDiagram}}

\figure{
\psfig{figure=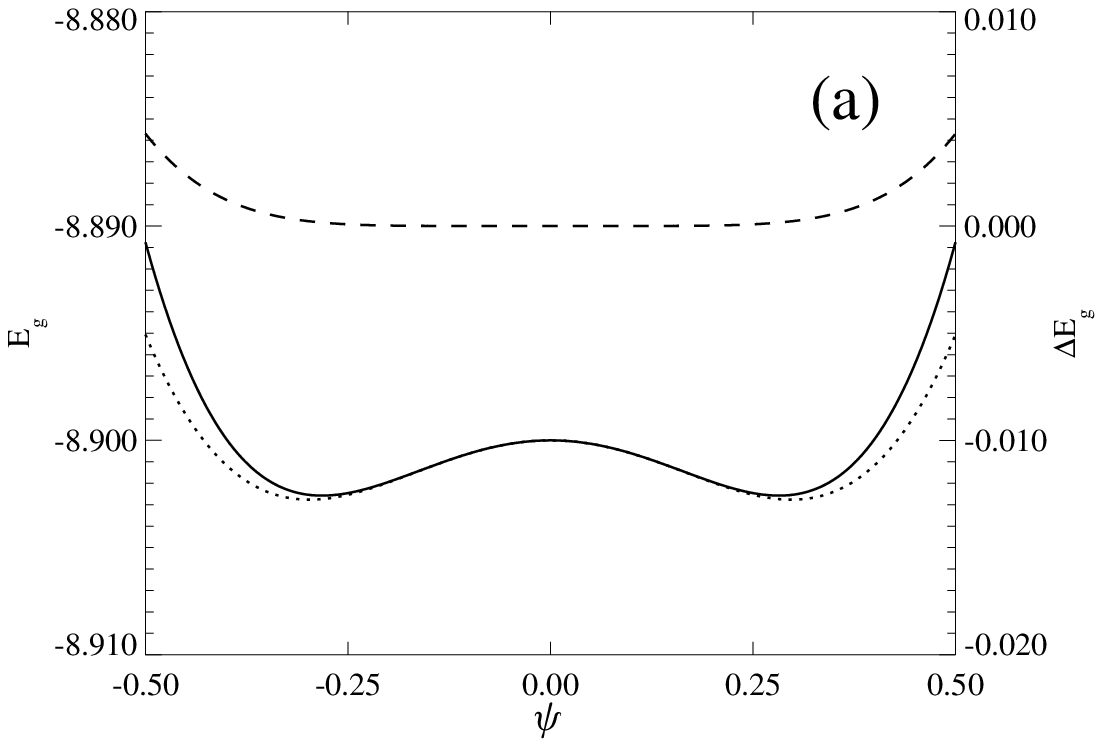,width=8cm}
\psfig{figure=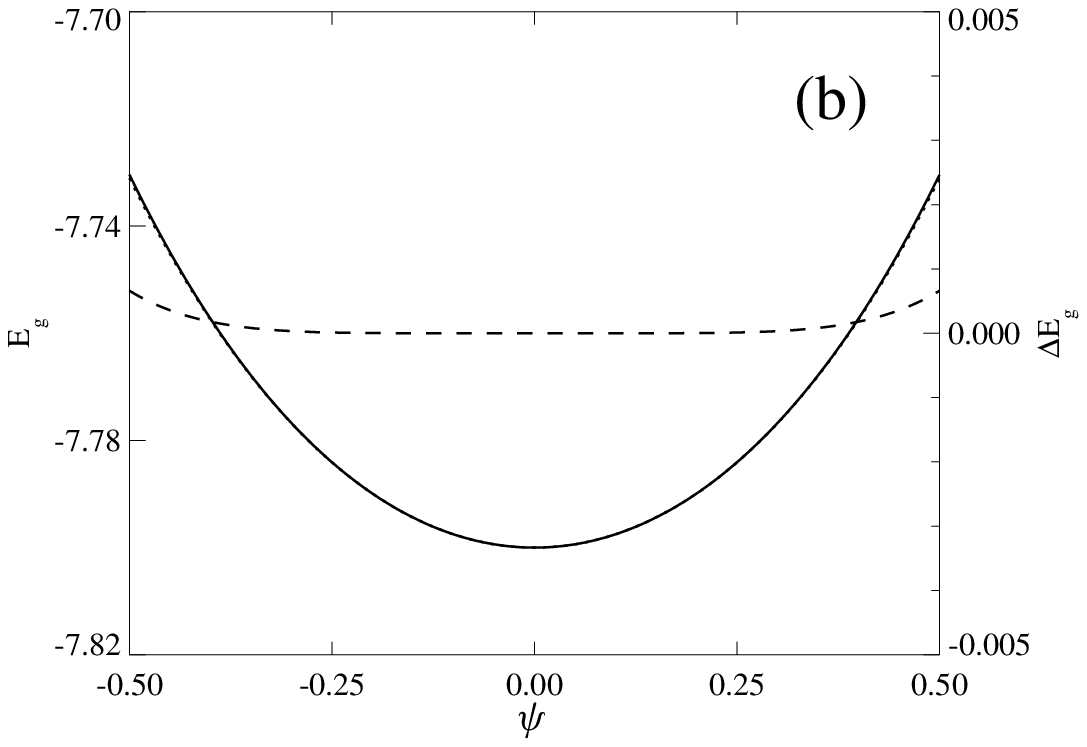,width=8cm}
\caption{\narrowtext Ground-state energy as a function of
$\psi$ for (a) ${\bar U}=11$ and
${\bar \mu}=8.9$ and for (b) ${\bar U}=11$ and ${\bar \mu}=7.8$.
The solid line represents fourth-order perturbation theory whereas the
dotted line represents a numerical diagonalization of the effective
hamiltonian. The dashed line is the difference between the
two (scale on the right).}
\label{Energieverg}}

\figure{
\psfig{figure=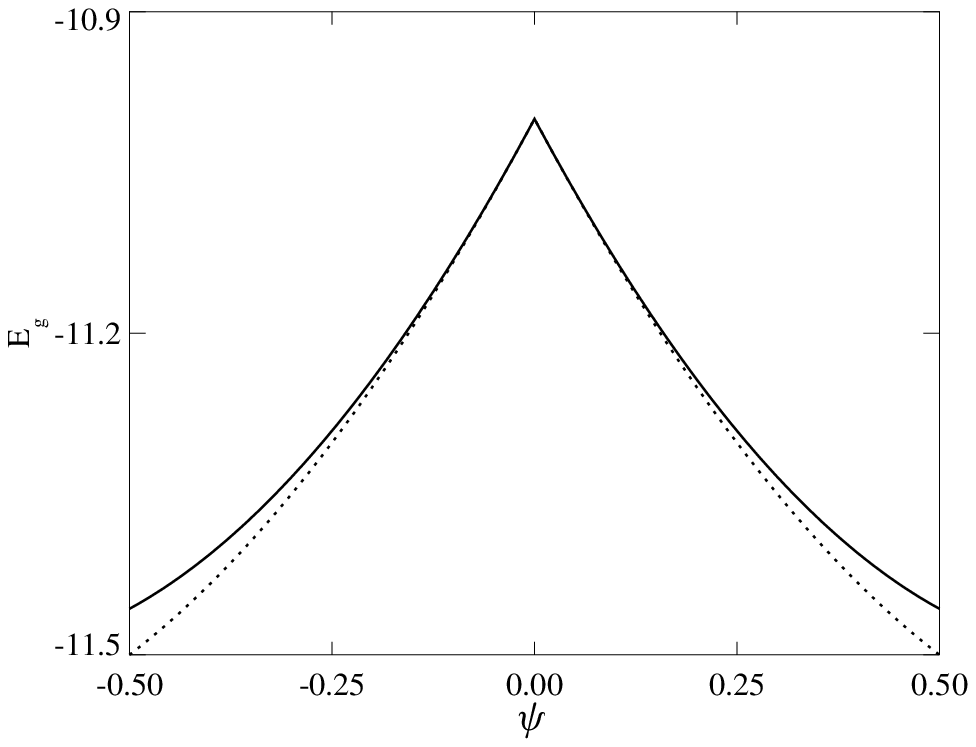,width=8cm}
\caption{\narrowtext Ground-state energy as a function of $\psi$ 
for ${\bar U}={\bar \mu}=11$, as obtained from first-order perturbation 
theory (solid line) and from numerical diagonalization of the effective
hamiltonian (dotted line).}
\label{Energiecusp}}

\figure{
\psfig{figure=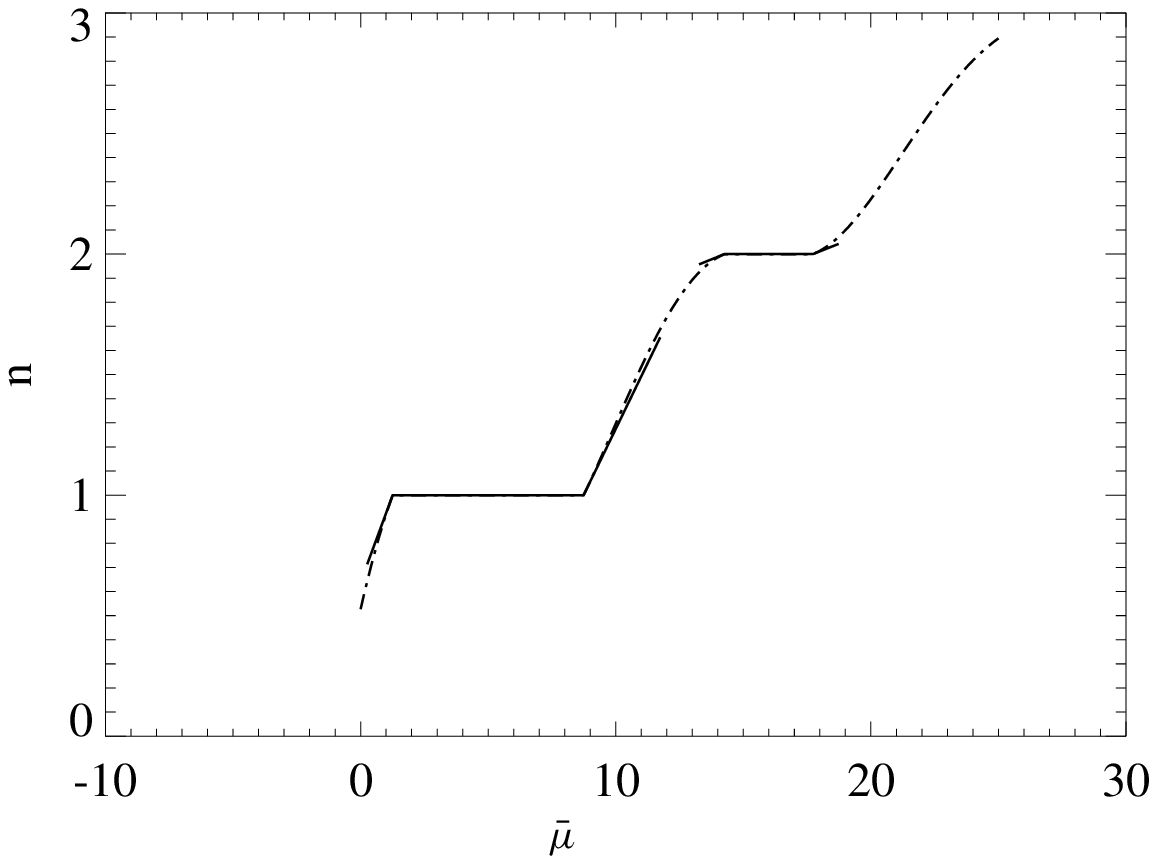,width=8cm}
\caption{\narrowtext The density as a function of chemical potential
${\bar\mu}=\mu/zt$ for an interaction strength of ${\bar
U}=U/zt=11$, i.e., along the dashed line in Fig.\ \ref{PhaseDiagram}.}
\label{DensityProfile}}

\figure{
\psfig{figure=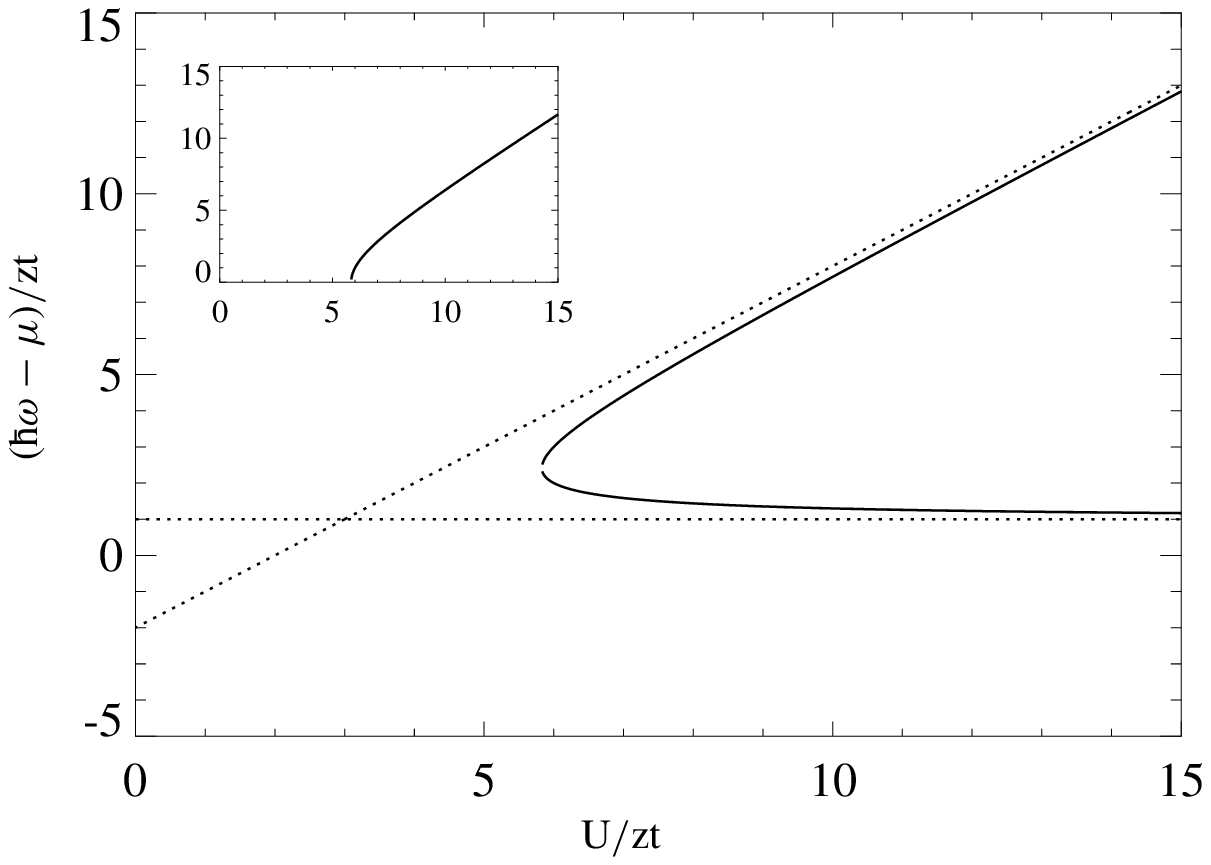,width=8cm}
\caption{\narrowtext The quasiparticle and the quasihole energy 
for ${\bf k}=0$  in the $g=1$ insulator lobe. The dotted lines are the 
asymptotes of the curves. The inset shows the resulting first-order 
approximation to the dispersion of the density fluctuations.}
\label{PHDispersie}}

\figure{
\psfig{figure=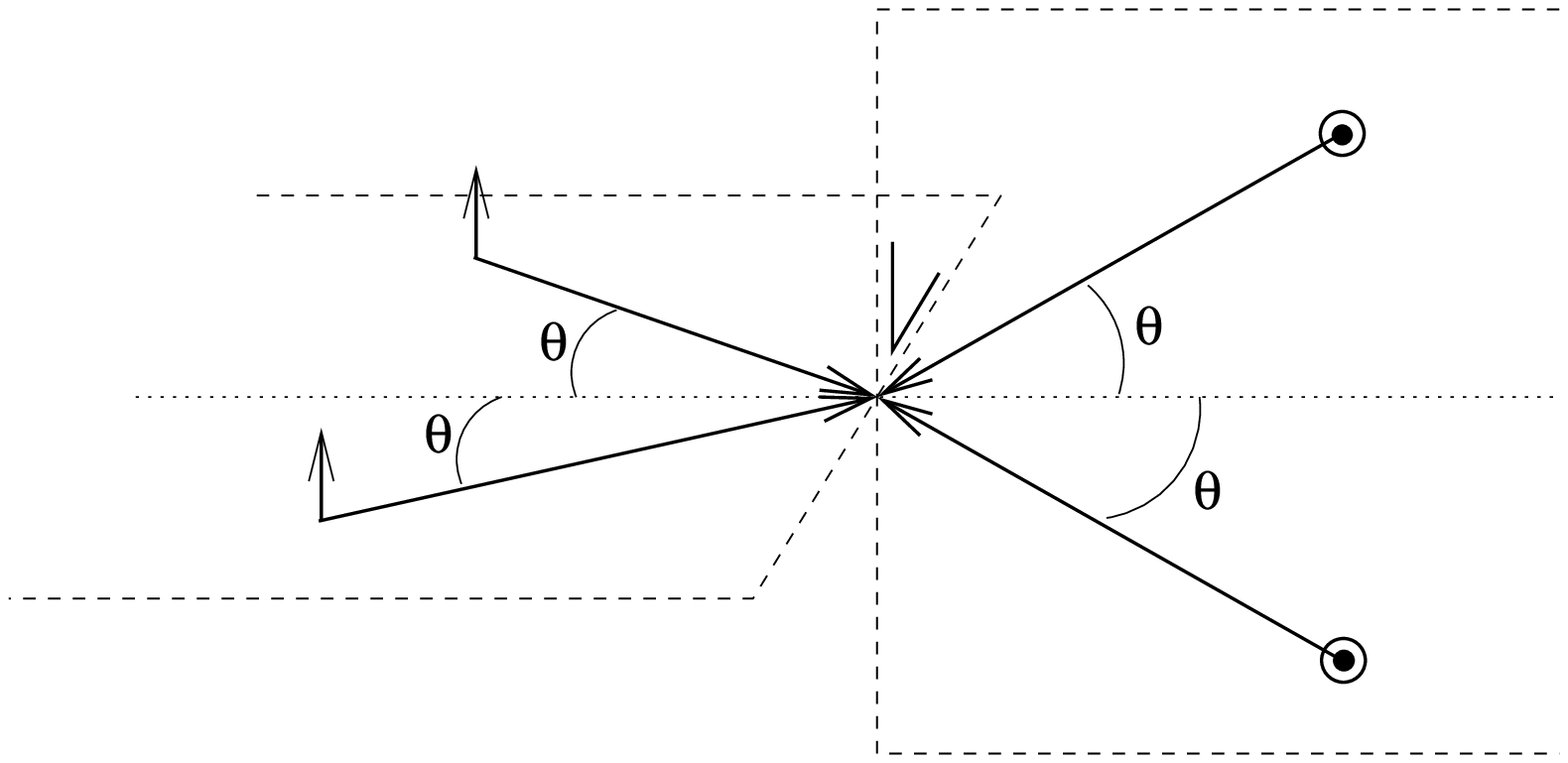,width=8cm,angle=-90}
\caption{\narrowtext Laser beam configuration for a 
three-dimensional optical lattice.}
\label{Configuratie}}

\figure{
\psfig{figure=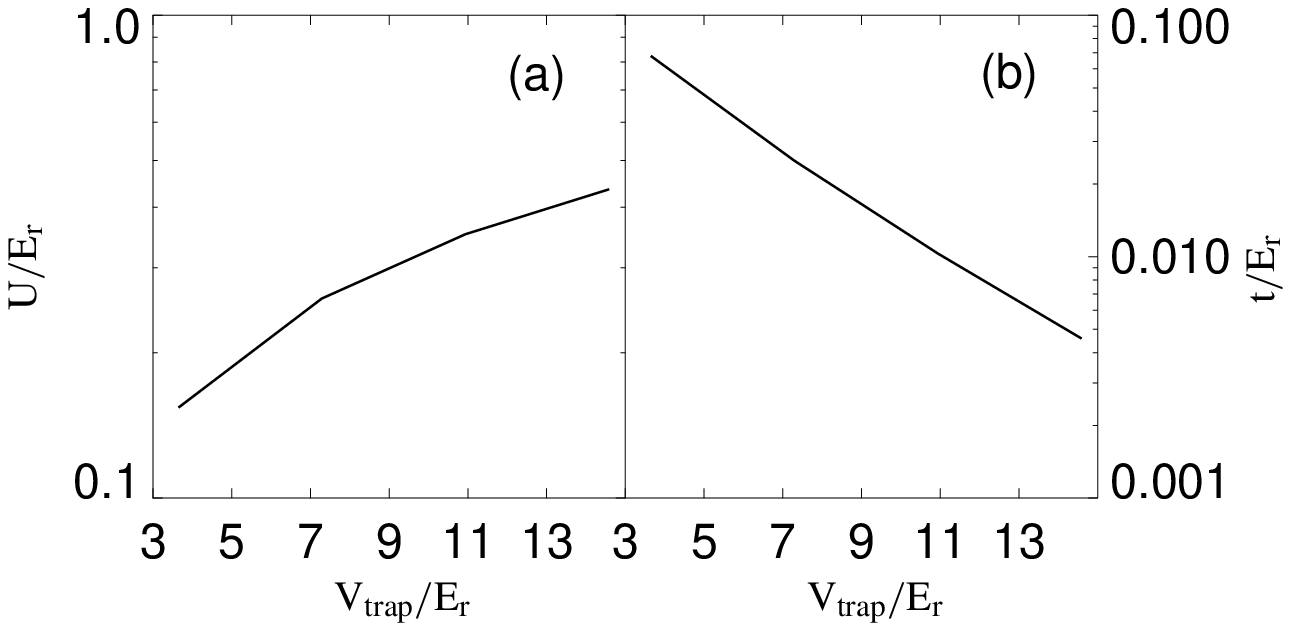,width=8cm}
\caption{\narrowtext
Plot of (a) $U$ and (b) $t$ as a function of the trap depth.
All quantities are in units of the recoil energy $E_r$.
}
\label{Parameterplaatjes}}

\figure{
\psfig{figure=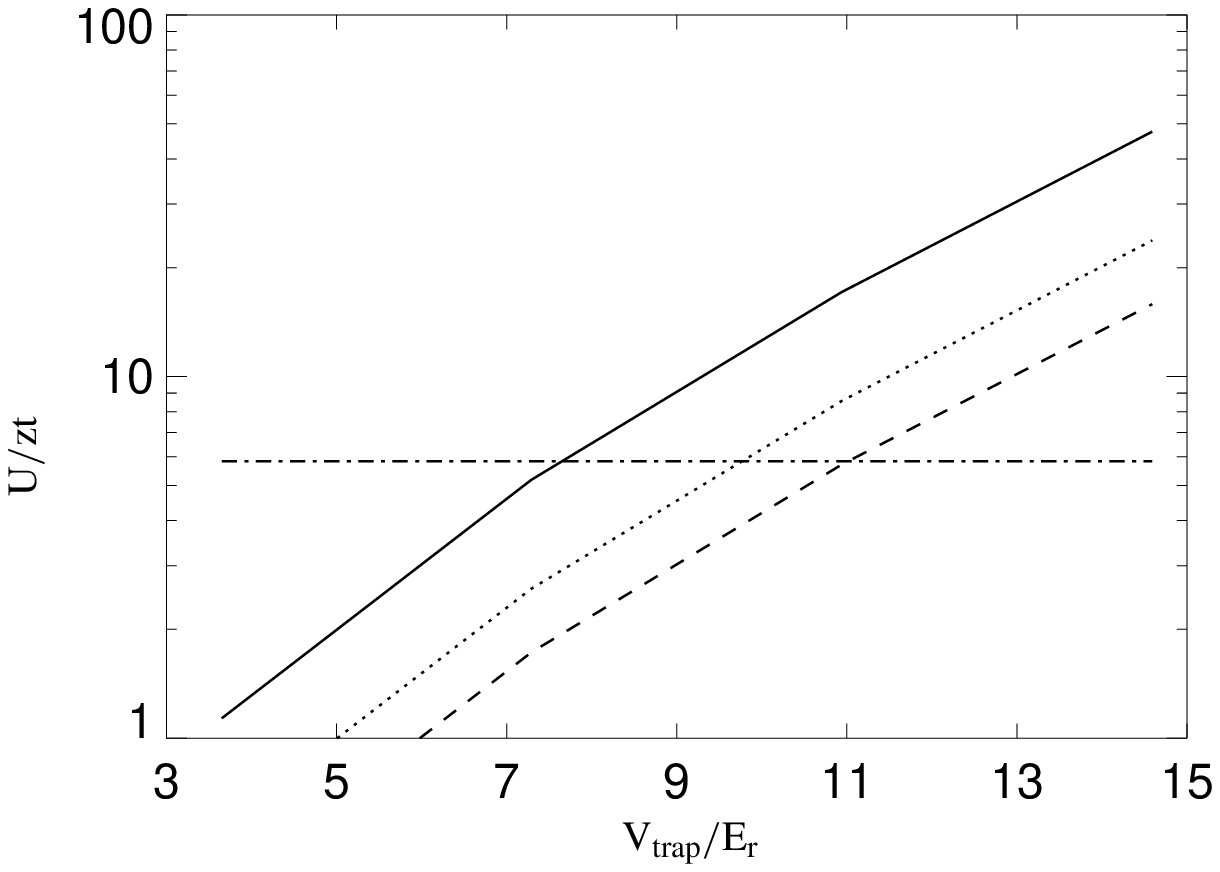,width=8cm}
\caption{\narrowtext 
The dimensionless parameter $U/zt$ plotted as a function of the trap depth
for $1$ (solid line), $2$ (dotted line) and $3$ dimensions (dashed
line). The dash-dotted line is the critical value ${\bar U}=5.83$.
}
\label{Uovertplaatje}}

\end{multicols}
\end{document}